%% file: main.tex
\documentclass[conference]{IEEEtran}
\IEEEoverridecommandlockouts
\pdfoutput=1 
\usepackage{cite}
\usepackage{amsmath,amssymb,amsfonts}
\usepackage{algorithmic}
\usepackage{graphicx}
\usepackage{textcomp}
\usepackage{xcolor}
\def\BibTeX{{\rm B\kern-.05em{\sc i\kern-.025em b}\kern-.08em
    T\kern-.1667em\lower.7ex\hbox{E}\kern-.125emX}}

\usepackage{float}
\usepackage{amsthm}
\usepackage{enumitem}
\usepackage[ruled,linesnumbered]{algorithm2e}
\usepackage{float}
\usepackage[normalem]{ulem}
\usepackage{subcaption}

\newtheorem{num1}{Theorem}
\newtheorem{num2}{Assumption}
\newtheorem{num3}{Definition}
\newtheorem{theorem}[num1]{Theorem}
\newtheorem{assumption}[num2]{Assumption}
\newtheorem{definition}[num3]{Definition}
\newcommand\hpm{\mathcal{M}}
\newcommand\hpmexo{\mathcal{U}_{\text{exo}}}
\newcommand\hpmendo{\mathcal{U}_{\text{endo}}}
\newcommand\hpmvarspace{\mathcal{V}}
\newcommand\hpmedges{\mathcal{E}}
\newcommand\hpmproperty{\psi}
\renewcommand\vec[1]{\mathbf{#1}}
\newcommand\hpmcontext{\mathbf{v}_{\text{cxt}}}
\newcommand\sys{\mathcal{S}}

\newcommand\func{f}

\newcommand\solfunc{f^*}
\newcommand\inspace{\mathcal{X}}
\newcommand\outspace{\mathcal{Y}}
\newcommand\property{\varphi}

\newcommand\hpmendoio{\mathcal{U}_{\text{io}}}

\newcommand\valassignio{\vec{v}_{\text{io}}}
\newcommand\nodediff[1]{\operatorname{diff}_{#1}}
\newcommand\nodediffio{\operatorname{diff}_{\hpmendoio}}
\newcommand\reconfuncs{\mathcal{F}_r}

\newcommand\hpmvarspaceendoio{\mathcal{V}_{\text{io}}}

\newcommand\simu{\textsc{simulate}}
\newcommand\tabu{\textsc{tabulate}}
\newcommand\discretize{\textsc{discretize}}
\newcommand\recon{\textsc{recon}}
\newcommand\encode{\textsc{encode}}
\newcommand\decode{\textsc{decode}}
\newcommand\comp{\mathcal{C}}

\newtheorem{innercustomthm}{Theorem}
\newenvironment{customthm}[1]
  {\renewcommand\theinnercustomthm{#1}\innercustomthm}
  {\endinnercustomthm}

\newcommand\IE[1]{\textcolor{blue}{#1}} 

\newcommand\edit[1]{#1}
\newcommand\editn[1]{#1}
\newcommand\editcr[1]{#1}

\begin{document}

\title{Causal Repair of Learning-Enabled\\ Cyber-Physical Systems}

\makeatletter
\newcommand{\linebreakand}{%
  \end{@IEEEauthorhalign}
  \hfill\mbox{}\par
\mbox{}\hfill
\begin{@IEEEauthorhalign}
}

\author{\IEEEauthorblockN{Pengyuan Lu}
\IEEEauthorblockA{\textit{Computer and Information Science} \\
\textit{University of Pennsylvania}\\
Philadelphia, PA, USA \\
pelu@seas.upenn.edu}
\and
\IEEEauthorblockN{Ivan Ruchkin}
\IEEEauthorblockA{\textit{Electrical and Computer Engineering} \\
\textit{University of Florida}\\
Gainesville, FL, USA \\
iruchkin@ece.ufl.edu }
\and
\IEEEauthorblockN{Matthew Cleaveland}
\IEEEauthorblockA{\textit{Computer and Information Science} \\
\textit{University of Pennsylvania}\\
Philadelphia, PA, USA \\
mcleav@seas.upenn.edu}
\linebreakand
\IEEEauthorblockN{Oleg Sokolsky}
\IEEEauthorblockA{\textit{Computer and Information Science} \\
\textit{University of Pennsylvania}\\
Philadelphia, PA, USA \\
sokolsky@cis.upenn.edu}
\and
\IEEEauthorblockN{Insup Lee}
\IEEEauthorblockA{\textit{Computer and Information Science} \\
\textit{University of Pennsylvania}\\
Philadelphia, PA, USA \\
lee@cis.upenn.edu}
}

\maketitle

\begin{abstract}
Models of actual causality leverage domain knowledge to generate convincing diagnoses of events that caused an outcome. It is promising to apply these models to diagnose and repair run-time property violations in cyber-physical systems (CPS) with learning-enabled components (LEC). However, given the high diversity and complexity of LECs, it is challenging to encode domain knowledge (e.g., the CPS dynamics) in a scalable actual causality model that could generate useful repair suggestions. In this paper, we focus causal diagnosis on the input/output behaviors of LECs. Specifically, we aim to identify which subset of I/O behaviors of the LEC is an actual cause for a property violation. An important by-product is a counterfactual version of the LEC that repairs the run-time property by fixing the identified problematic behaviors. Based on this insights, we design a two-step diagnostic pipeline: (1) construct and Halpern-Pearl causality model that reflects the dependency of property outcome on the component's I/O behaviors, and (2) perform a search for an actual cause and corresponding repair on the model. We prove that our pipeline has the following guarantee: if an actual cause is found, the system is guaranteed to be repaired; otherwise, we have high probabilistic confidence that the LEC under analysis did not cause the property violation. We demonstrate that our approach successfully repairs learned controllers on a standard OpenAI Gym benchmark.
\end{abstract}


\input{sections/1_intro}
\input{sections/2_background}

\input{sections/3_problem}
\input{sections/4_hp_model}
\input{sections/5_diagnosis}

\input{sections/6_experiments}
\input{sections/7_discussion}
\input{sections/8_conclusion}
\input{sections/acknowledgement}

\bibliographystyle{IEEEtran}
\input{sections/references}

\newpage
\appendices 
\input{sections/Appendix_A_proofs}

\end{document}

%% file: sections/1_intro.tex
\section{Introduction}
\label{sec:introduction}

When a person's leg hurts, they seek detailed diagnosis from a physician regarding the pain's cause, which would lead to an effective intervention \edit{as a ``repair''}. Similarly, when a closed-loop cyber-physical system (CPS) violates a desirable property at run time, the violation needs diagnosis --- a procedure that identifies the cause for the violation, and by fixing the identified cause, the CPS can be repaired. Traditionally, researchers conduct this kind of analysis from statistical inference on observations~
\cite{wang2013causality, schumann2015r2u2}
. However, statistical diagnosis is prone to mistaking correlation for causation: when a student always wears a green jacket and fails several exams, such algorithms are likely to conclude it is the green jacket's fault due to the perfect correlation. Therefore, in this paper, we focus on stronger causal reasoning and repair on CPS failures.

\looseness=-1
Researchers have explored the concept of \textit{actual causality} that leverages domain knowledge to produce well-defined and convincing causal explanations. Informally, an actual cause for an outcome is a minimal set of variable assignments, which represent an event, that changes the outcome if assigned some counterfactual values. Finding an actual cause requires the construction of an actual causality model, such as a Halpern-Pearl model, which rigorously defines actual causes for events and precisely assigns responsibility and blame~\cite{halpern2005causes, halpern2016actual, chockler2004responsibility}. With these models, we can encode common knowledge that, for instance, the student's bad grade can be either due to not studying hard enough or misunderstanding some concepts in class. 


\edit{Although} actual causality is \edit{promising} for analyzing and fixing CPS failures, causal analysis and repair have been complicated by the growing popularity of learning-enabled components (LECs) \cite{held2016learning, hartsell2019model, tuncali2018reasoning}. First, LECs usually consist of numerous internal continuous parameters, such as  weights and biases in deep neural networks. Second, LECs take a large diversity of forms, from basic statistical models such as linear regressors and support vector machines to complex deep architectures, lacking a shared struture of a parameter template. 
Furthermore, sometimes the internal structures of LECs are black-boxes protected as intellectual properties, and are invisible to testing engineers. Under these scenarios, it is hard to build a causal model based on the internal information flow of these components.
Related to diagnosis and repair are the efforts in explainable AI~\cite{chakraborti2017plan, krarup2019model} and formal methods~\cite{raman2012explaining, raman2013towards}, where researchers build frameworks to explain behaviors of learned agents. However, to our knowledge, these strands of work have yet to connect to actual causality.

\edit{Since the internal parameters of LECs are hard to analyze, we take a step back and analyze the LEC I/O behavior instead.} Our intention in this paper is to leverage actual causality to efficiently identify the granular I/O behaviors of a suspected LEC for a run-time property violation. We intend to identify these behaviors in the form of fine-grained input-output mappings, so that a precise repair can be made for CPS to satisfy the desired property. 

\editn{To achieve this goal, we first construct an Halpern-Pearl (HP) causality model to encode the dependency of property outcome on the suspected LEC's behaviors. Then, we design an algorithm to search for an actual cause and corresponding repair on this HP model.}
This algorithm either concludes that an actual cause does not exist \editn{within the LEC's behaviors} with high confidence 
, or outputs the set of behaviors that are causing the violation, along with a counterfactual repair. Experiments on the mountain car, a standard OpenAI Gym benchmark~\cite{brockman2016openai}, \editn{demonstrates the capability of our method.} 

To summarize, our contributions are: 
\begin{enumerate}
    \item A novel use of HP causality model to identify problematic LEC I/O behaviors that cause a run-time property violation, and produce repair suggestions, without touching the component's internal information flow, and
    \item Experiments on an OpenAI Gym benchmark that show our solution's utility for learning-enabled CPS repair.
\end{enumerate}

%% file: sections/2_background.tex
\section{Background and Related Work}
\label{sec:background}

\subsection{Learning-enabled Components (LEC)}
\label{subsec:background_lec}

\looseness=-1
\edit{Learning-enabled components (LECs) are functional components of larger systems that are learned from data, either offline or online.} 
One example is the perception unit, with neural networks 
able to complete difficult vision tasks that generally cannot be accomplished by traditional first-principles algorithms, such as tracking moving objects at 100 fps~\cite{held2016learning}. Moreover, deep reinforcement learning has provided useful control policies on various tasks as a substitute for conventional controllers~\cite{qi2019deep, cao2021deep}. Due to LECs' performance, the state-of-the-art in CPS has a growing enthusiasm of these statistically generated agents, with automated design tools built for CPS with LECs~\cite{hartsell2019model}, as well as analysis on their assurance on run-time properties such as safety~\cite{tuncali2018reasoning}. Consequently, we need to consider the presence of LECs when diagnosing faults in CPS.

\subsection{Repair}
\label{subsec:background_repair}

A repair is a procedure to change or replace parts of a system to achieve a desirable performance, of which the system previously fell short. This term has been widely used in traditional embedded systems. For example, researchers have studied repair on hardware components such as DRAM ~\cite{mcconnell2001test, zorian2002embedded} \editcr{or noisy sensors~\cite{cazes2020model}}. Also, repair has been performed on software; for example, assembly program can be transformed to decrease resource consumption below a threshold~\cite{schulte2013automated}.

In modern CPS with LEC, the topic of repair interests many researchers. Learned agents, such as deep neural networks, may lead the system to unsafe states~\cite{cruz2021safe} or unplanned paths~\cite{peddi2021interpretable}. Therefore, repair is necessary to recover desirable behaviors. \editcr{Repair of neural networks is an active area of research~\cite{islam2020repairing,majd2021local}.
For instance, network parameters can be repaired by search algorithms \cite{sohn2019search} or constraint solvers \cite{usman2021nn}.
More recently, researchers have studied provable repairs on deep neural networks, with guaranteed satisfaction of a given property after the repair~\cite{lin2020art,cohen2022automated,fu2022reglo}. } 


\editcr{Compared to the existing work on repair, we focus on the causal relationships between the repairable elements and the execution outcome. That is, the neural network parameters to be repaired should be the ones that caused the system's failure.}


\subsection{Actual Causality and Halpern-Pearl Models}
\label{subsec:background_actual_causality}

\edit{Below} we rephrase Halpern and Pearl's definition of actual causality \cite{halpern2005causes, halpern2016actual}. 
An actual causality model, or a Halpern-Pearl (HP) model is a recursive structure, i.e., a directed acyclic graph,
such that every node represents either an \textit{exogenous variable}, whose value is determined by factors outside this model, or an \textit{endogenous variable}, whose value is determined by other variables in this model. The edges represent dependencies: an exogenous node has only outgoing edges but no incoming edges, while an endogenous node may have both. Every endogenous node is equipped with a function, which defines how the node's value is computed from other nodes. In other words, this function defines the incoming edges to the node. Formally,
\begin{enumerate}
    \item An \emph{HP model} is a tuple $\hpm = (\hpmendo, \hpmexo, \hpmvarspace, \hpmedges)$, \editn{where $\hpmendo$ and $\hpmexo$ are finite sets of endogenous and exogenous variables, respectively, and for each variable $u \in \hpmendo \cup \hpmexo$, $\hpmvarspace(u)$ defines a non-empty and potentially infinite set of values that $u$ can take.}
    \item $\hpmedges$ is a \edit{set of edges, associated with dependency equations,} that defines how the value of each endogenous node $u$ is computed. I.e., for each $e_u \in \hpmedges$, 
    \[e_u: \prod_{u' \in \hpmendo \setminus \{u\}} \hpmvarspace(u') \times \prod_{u'' \in \hpmexo} \hpmvarspace(u'') \mapsto \hpmvarspace(u).\] 
\end{enumerate}

To define an actual cause in an HP model $\hpm$, \edit{we introduce the following notation:}
\begin{enumerate}
    \item An \textit{assignment} of a variable $u \in \hpmendo \cup \hpmexo$ is denoted as $u := v$, for some $v \in \hpmvarspace(u)$. Assignments of multiple variables are denoted in vector form $\vec{u} := \vec{v}$, with $\vec{u} = [u_1, u_2, \dots]$, $\vec{v} = [v_1, v_2, \dots]$. 
    This assignment means a conjunction $(u_1:=v_1) \land (u_2 := v_2) \land \dots$.
    \item A \textit{property} $\hpmproperty$ is a Boolean function of endogenous variable assignments, e.g., $\hpmproperty = (u_1 := v_1) \land (u_2 := v_2) \lor (u_3:=v_3)$. The satisfaction relation $(\hpm, \hpmcontext) \models \hpmproperty$ denotes that a property holds on $\hpm$ given \editn{exogenous nodes assigned with $\hpmcontext$}.
    \item A \textit{counterfactual} $\vec{v}'$ (with respect to the ``factual'' $\vec{v}$) is an alternative value assignment on some endogenous variables. The values of dependent nodes in $\vec{v}'$ may be different from  $\vec{v}$ in accordance with to $\hpmedges$. A property $\hpmproperty$ holds on a counterfactual that replaces the factual values $\vec{v}$ with $\vec{v}'$ on variables $\vec{u}$ is denoted as $(\hpm, \hpmcontext) \models [\vec{u} \leftarrow \vec{v}']\hpmproperty$, or simply $[\vec{u} \leftarrow \vec{v}']\hpmproperty$.
\end{enumerate}

Then, in the above terms, the Halpern and Pearl definition of an actual cause is phrased as follows.

\begin{definition}[Actual Cause]
\label{def:actual_cause}
\editn{On an HP model $\hpm$ with exogenous node values $\hpmcontext$, the assignment on a set of endogenous variables $\vec{u}:=\vec{v}$ is an actual cause of $\hpmproperty$ iff the following conditions hold:}
\begin{enumerate}
    \item AC1: $(\hpm, \hpmcontext) \models (\vec{u}:=\vec{v}) \land \hpmproperty$
    \item AC2: $\exists$ partition $\hpmendo = \vec{u} \cup \vec{u}_1 \cup \vec{u}_2$. Denote the factual values of $\vec{u}_1$ and $\vec{u}_2$ as $\vec{v}_1$ and $\vec{v}_2$, respectively. Then, $\exists \vec{v}', \vec{v}_1'$, such that\\
    (a) $[\vec{u} \leftarrow \vec{v}', \vec{u}_1 \leftarrow \vec{v}_1']\neg\hpmproperty$\\
    (b) $[\vec{u} \leftarrow \vec{v}, \vec{u}_1 \leftarrow \vec{v}_1', \vec{u}_2^* \leftarrow \vec{v}_2^*]\hpmproperty$, for any $\vec{u}_2^* \subseteq \vec{u}_2$ and $\vec{v}_2^*$ is the original factual value of $\vec{u}_2^*$ (a subvector of $\vec{v}_2$).
    \item AC3: $\nexists \vec{u}\IE{'} \subset \vec{u}$ that satisfies AC1 and AC2.
\end{enumerate}
\end{definition}

Condition AC1 ensures the suspected actual cause and outcome are factual. Then, AC2(a) ensures a sufficient counterfactual, that switching the suspect $\vec{u}$ and a circumstance $\vec{u}_1$ to that counterfactual assignment guarantees a flipped outcome $\neg\hpmproperty$, and AC2(b) states that switching the circumstance alone does not change the outcome - as long as the suspect remains the factual value assignments. Finally, AC3 guarantees minimality of the actual cause. Detailed explanation for this definition can be found in the original publications \cite{halpern2005causes, halpern2016actual}.

Researchers have applied actual causality and HP model to CPS diagnosis. For instance, Ibrahim et al. have designed a SAT solver to practically compute actual causes to explain undesirable CPS behaviors \cite{ibrahim2019efficient, ibrahim2019practical}. However, the solver is restricted to finite $\hpmvarspace(u)$ for variables \edit{and their HP model design only captures discrete events like "there exists a Byzantine fault" or "the system is on autopilot mode".} 
\editn{Unfortunately, this design does not extend to diagnosis of LECs, which generally have continuous value spaces for I/O and internal variables, and their internal information flows are not interpretable.}

%% file: sections/3_problem.tex
\section{Problem Formulation}
\label{sec:problem}

\subsection{System Setting}
\label{subsec:sys_setting}

We formalize \edit{the system setting as follows}. We have an \edit{exact} dynamical system model $\sys$, which describes how the components of the agent interact with each other, as well as how the agent interacts with the environment. This closed-loop model is assumed to be deterministic. 
In other words, \edit{the agent's trajectory} only depends on the initial state, the environment dynamics, and the component designs. These assumptions are often satisfied in model-based CPS engineering and can be relaxed in future research.

The system aims to satisfy a given specification/property $\property$ that takes Boolean values (true/satisfied/1, false/violated/0), e.g., \editn{the linear temporal logic (LTL)~\cite{pnueli1977temporal} or signal temporal logic (STL)~\cite{maler2004monitoring} formula}. \edit{In this paper, we focus on STL.}

\editn{We observe a trajectory on a given initial state $s_0$, where the system failed to satisfy $\property$ and we suspect one of its components $\comp$, such as a controller, is at fault for the violation.} 
\edit{The component $\comp$'s} internal design is invisible to us, but we \edit{can observe its input/output (I/O) behavior, which we denote by function} $\func: \inspace \mapsto \outspace$, with its domain and codomain being \textit{continuous and bounded} metric spaces $(\inspace, d_\inspace)$ and $(\outspace, d_\outspace)$, respectively.
Therefore, we can replace this \edit{behavior} by any \textit{counterfactual} $\func': \inspace \mapsto \outspace$. We denote the counterfactual system that uses $\func'$ in place of $\func$, with everything else remaining the same, as $\sys(\func')$. We use $\sys(\func') \models \property$ and $\sys(\func') \not\models \property$ to denote \edit{that} the property will be satisfied or violated under $\sys(\func')$. \edit{With} this notation, the factual outcome is $\sys(\func) \not\models \property$.
\editn{
Next, we formalize the distance between two choices for the behaviors of component $\comp$.
\begin{definition} [Distance between I/O behaviors]
\label{def:component_distance}
For any two I/O behaviors $\func_1, \func_2: \inspace \mapsto \outspace$, we define distance $||\cdot||_{d_\func}$ as
\begin{equation}
    \label{eq:component_distance}
    ||\func_1 - \func_2||_{d_\func} = \max_{x \in \inspace} ||\func_1(x) - \func_2(x)||_{d_\outspace}
\end{equation}
\end{definition}
We then make the following assumptions.}

\editn{
\begin{assumption}
    \label{assum:simulator}
    We can check the outcome of $\property$ on the given initial state $s_0$ when substituting different $\func'$ in place of the component by calling a simulator,
    \begin{equation}
        \label{eq:simulator}
        \simu_{s_0}: (\inspace \mapsto \outspace)  \mapsto \{0, 1\},
    \end{equation}
    which encodes the knowledge of $\sys$ and $\property$. For simplicity, we assume a fixed $s_0$ in the remainder of this paper, and the simulator is denoted as simply $\simu$.
\end{assumption}
\begin{assumption}
    \label{assum:lipchitiz}
    The behavior $\func$ is Lipschitz-continuous, with an unknown Lipschitz constant. This is a common property of many types of learning models, such as neural networks.
\end{assumption}
\begin{assumption}
    \label{assum:robustness}
    The property outcome is robust against small changes from the factual I/O behavior $\func$, i.e., $\forall \func': \inspace \mapsto \outspace$,
    \begin{equation}
        \label{eq:robustness}
        ||\func - \func'||_{d_\func} \leq \epsilon \implies \simu(\func) = \simu(\func'),
    \end{equation}
    for some small $\epsilon > 0$.
\end{assumption}}

\editn{
We expect Assumption \ref{assum:robustness} to hold in most practical cases. Suppose the distance of the factual $\func$ to the decision boundary of the simulator outcome on I/O behaviors is $\delta$. If $\delta > 0$, there exists an arbitrarily small $\epsilon < \delta$ where the assumption holds. The only case that Assumption \ref{assum:robustness} does not hold is when $\delta = 0$, i.e., the factual behavior is right on the decision boundary, but this event would usually have a probability measure of $0$.
}

\subsection{Problem Statement}

\edit{In the above setting}, we want to identify a subset of I/O behaviors of the suspected component \edit{$\comp$} --- that is, a set of input-output tuples of $\func$ --- that indeed caused the property violation\edit{, as well as the counterfactual outputs on these inputs that can repair the system.}

\textbf{Main Problem.} Upon the observation of a property
violation \edit{on a runtime trace},
$\sys(\func) \not\models \property$, how \edit{can we use} HP causality to identify a subset of the suspected component \edit{$\comp$}'s I/O behaviors (modeled as $\func$) that caused this violation? 

\begin{enumerate}
    \item \textbf{Sub-problem 1.} \edit{Encode the dependency structure of $\comp$'s behaviors on $\property$ using an HP model.} 
    \item \textbf{Sub-problem 2.} On the encoded HP model, \edit{design a search algorithm for an actual cause}, such that, upon success, provide a repair suggestion in form of a counterfactual $\func^*: \inspace \mapsto \outspace$ that $\sys(\func^*) \models \property$. Upon failure, \edit{quantify the confidence that} the property violation is not \edit{caused by} 
    \edit{the I/Os of $\comp$}.
\end{enumerate}

\begin{figure}
    \centering
    \includegraphics[width=0.5\textwidth]{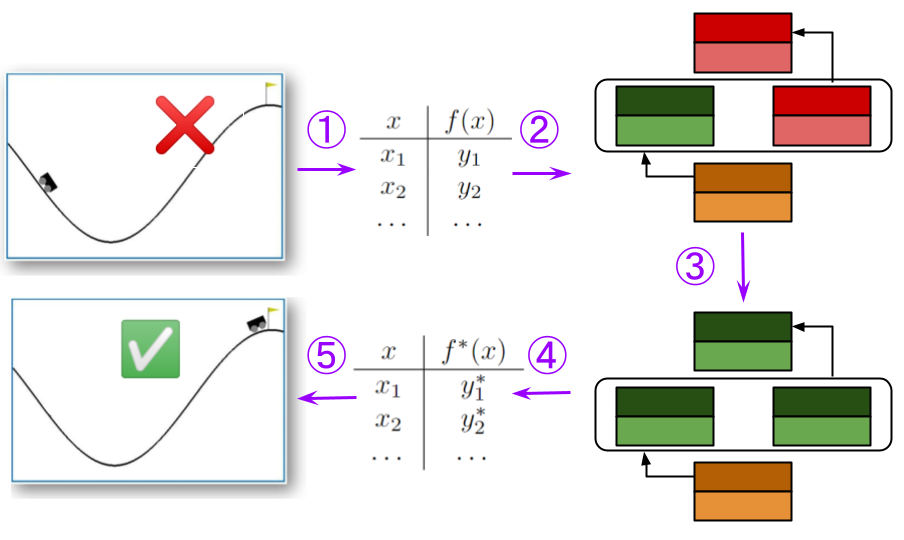}
    \caption{Workflow of our \edit{causal repair} approach\edit{, which constructs an HP model to encode the I/O behaviors of a suspected LEC and search for a repair by causal analysis.}}
    \label{fig:workflow}
\end{figure}

\edit{To solve these problems, we employ the following workflow, which is visualized in Figure~\ref{fig:workflow}: }\editn{(1) Extract the behaviors of $\comp$ as an I/O table. (2) Encode the dependency of the property outcome on the I/O behaviors with an HP model (Section \ref{sec:hp_model}). (3) Search for a counterfactual model value assignment, revealing an actual cause and a repair (Section \ref{sec:diagnosis}). (4) Decode the found assignment as a counterfactual component behavior. (5) Replace $\comp$ with an alternative component that performs this counterfactual behavior to repair the system.}


The entire workflow is implemented on an OpenAI Gym example in Section \ref{sec:expriments}.

%% file: sections/4_hp_model.tex
\section{Halpern-Pearl Model Design}
\label{sec:hp_model}

We first describe a naive way of encoding the dependency of $\property$ on component $\mathcal{C}$'s I/O behaviors in Section~\ref{subsec:infinite_hp_model}. This results in an HP model with an infinite number of nodes. To alleviate this issue, we describe a method for constructing an HP model with a finite number of nodes in Section~\ref{subsec:discrete_hp_model}. 

\editn{
To fulfill the minimality in AC3 of Definition~\ref{def:actual_cause}, we will need to distinguish which counterfactuals are closer to the factual behavior. This requires a  partial order on I/O behaviors that aligns with some partial order on the HP node values. Unfortunately, the naive finite HP model does not admit fine enough partial orders over its node values. Thus, we will refine our HP model to support suitable partial orders and make it amenable to the counterfactual search.}

\subsection{Infinite HP Model}
\label{subsec:infinite_hp_model}

\begin{figure}[t!]
    \centering
    \includegraphics[width=0.35\textwidth]{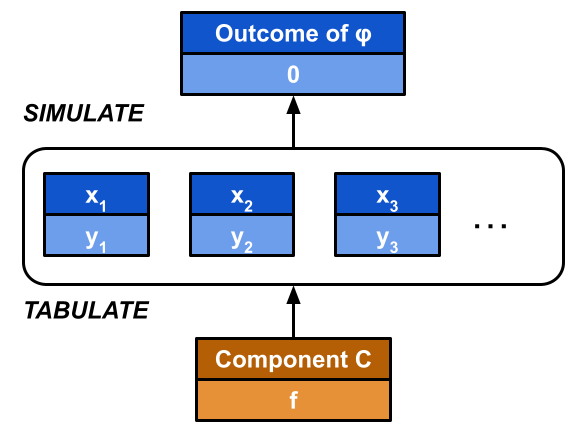}
    \caption{\edit{Infinite HP model.}}
    \label{fig:hp_model_ver_1}
\end{figure}

Intuitively, a naive way to build \edit{the} HP model \editn{to model the dependency of property outcome on I/Os of $\comp$}
is to model every input $x \in \inspace$ as an endogenous node, and its corresponding output $y \in \outspace$ as its value. \edit{We illustrate this model in Figure~\ref{fig:hp_model_ver_1} and refer to it as the \emph{infinite HP model}}.

This model encodes \editn{the component $\comp$'s implemented I/O behaviors as an exogenous node (yellow), and the behaviors for each input $x \in \inspace$, as well as the property outcome as endogenous nodes. Since we assume the full knowledge of the $\comp$'s I/Os, we can extract the input-output tuples by tabulating them. 
Then, based on Assumption \ref{assum:simulator}, the outcome can be obtained by calling $\simu$ on the I/Os. This design is illustrated in Figure~\ref{fig:hp_model_ver_1}. Each node has its variable name at the top and a value at the bottom.} 

Recall that the input space $\inspace$ is continuous. Therefore, \edit{an} obvious drawback of this \edit{infinite HP model} is that \edit{it has} infinitely many nodes for each $x \in \inspace$, and \edit{thus an} infinitely large search space for an actual cause given by Definition \ref{def:actual_cause}. 

\subsection{Discretized HP Model}
\label{subsec:discrete_hp_model}

\edit{To create a finite HP model, we partition} the input space $\inspace$ and output space $\outspace$ into sufficiently fine-grained, but finitely many \editn{cells}
$\inspace = \bigcup_{i=1}^m \inspace_i$ and $\outspace = \bigcup_{j=1}^n \outspace_j$, and then \edit{consider functions that map all $x \in \inspace_1$ to the center of one $\outspace_j$.}
The partitioning procedure is detailed in Algorithm~\ref{alg:io_partition}.

\begin{algorithm}[t]
\caption{\editn{Discretize HP Model}} \label{alg:io_partition}
\SetKwInOut{Input}{Input}
\SetKwInOut{Output}{Output}
\SetKwRepeat{Do}{do}{while}
\Input{\editn{Factual behavior $\func: \inspace \mapsto \outspace$, continuous bounded spaces $\inspace$ and $\outspace$, initial cell width $\Delta_x^ {\text{init}}$, $\Delta_y^ {\text{init}}$}}
\Output{Partitioned cells $\inspace_1, \dots, \inspace_m$, $\outspace_1, \dots, \outspace_n$, \edit{a map $g: \{1, \dots, m\} \mapsto \{1, \dots, n\}$ from input cells to output cells}}
$\Delta_x \gets \Delta_x^ {\text{init}}$\;
$\Delta_y \gets \Delta_y^ {\text{init}}$\;
\Do{$\simu(\func_r) \neq \simu(\func)$}{
    Partition $\outspace$ with cell width $\Delta_y$\;
    $\Delta_y \gets \Delta_y / 2$\;
    \Do{$\exists \inspace_i$ \text{that cannot be completely mapped} \text{into some} $\outspace_j$}{
        Partition $\inspace$ with cell width $\Delta_x$\;
        $\Delta_x \gets \Delta_x / 2$\;
    }
   \edit{$g \gets$ the current mapping between cells\;
   $\func_r \gets \recon(g)$\;}
}
\end{algorithm}

Algorithm \ref{alg:io_partition} splits the input space and output space into hypercube cells. The cell size keeps shrinking until the two conditions, respectively in line 10 and line \edit{13} are met. \edit{We define the reconstruction method $\recon$ at line 12 as follows.
\begin{equation*}
    \recon(g) = \func_r, \text{ where } \forall x \in \inspace_i, \func_r(x) = \text{center}(\outspace_j), j = g(i)
\end{equation*}
That is, it reconstructs \editn{an I/O behavior} from the discrete map $g$\edit{, which approximates the factual I/O behaviors as a mapping among cells,}
by mapping every $x$ in an input cell to the center of the corresponding output cell.}
This algorithm is guaranteed to terminate \edit{by the following theorem.}

\begin{theorem}[Termination of Discretization]
    Algorithm \ref{alg:io_partition} is guaranteed to terminate.
\end{theorem}
\editcr{For the proof, please refer to Appendix \ref{sec:appendix_proofs}. Proofs of later theorems can also be found there.} 


Upon the termination of Algorithm \ref{alg:io_partition}, we obtain a function \edit{$\func_r = \recon(g)$} that represents $\func$ by (1) having arbitrarily close outputs on the same inputs 
and (2) having the same property outcome. Then, our causal analysis is established on the family of functions like $\func_r$, i.e.,
\begin{definition}[Representative Component \editn{Behavior} Space]
\label{def:representative_space}
\edit{Given the cell partition $\inspace = \bigcup_{i}^{m}\inspace_i$ and $\outspace = \bigcup_{j}^{n}\outspace_j$, } we define the \emph{representative component \edit{behavior} space} as a finite subset of $(\inspace \mapsto \outspace)$ as
\begin{equation}
    \label{eq:representative_space}
    \reconfuncs = \{\func' \in (\inspace \mapsto \outspace) \mid \forall i, \forall x \in \inspace_i, \exists j, \func'(x) = \operatorname{center}(\outspace_j)\}
\end{equation}
\end{definition}
\editn{Notice that if we wrap up the generation of $g$ from $\func$ in Algorithm \ref{alg:io_partition} as a method $\discretize$, the two methods $\discretize$ and $\recon$ are inverse to each other if the behaviors are restricted within $\reconfuncs$.}

\begin{figure}
    \centering
    \includegraphics[width=0.4\textwidth]{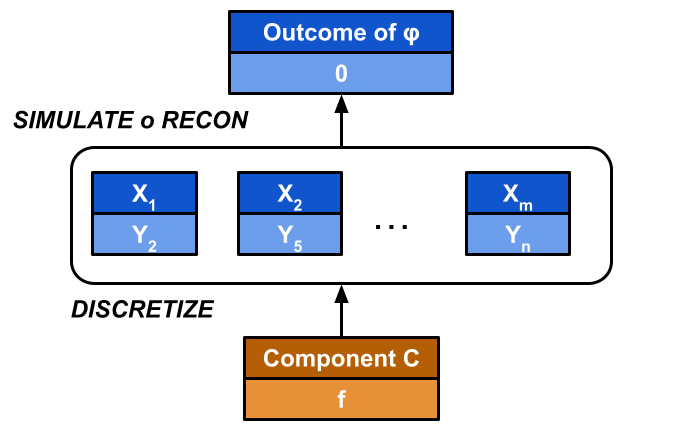}
    \caption{\edit{Discretized HP model.} We illustrate this with an example value assignment of nodes.}
    \label{fig:hp_model_ver_2}
\end{figure}

With this shrunken space of \editn{behavior} choices, we modify our HP model to \edit{the next version, called the \emph{discretized HP model}}, as in Figure \ref{fig:hp_model_ver_2}. Here, we have one endogenous node per input cell, and its value is the mapped output cell. This HP model has $m+2$ 
nodes and can express every \editn{behavior} choice in $\reconfuncs$. 
\edit{Notice that in place of $\tabu$, we now have $\discretize$ since the nodes are now representing the discrete map $g$, and the $\simu$ method requires $\recon$ on the discrete map first.}

\editn{Generally, we want to pick a counterfactual for $\comp$ that repairs the system $\sys$, i.e., flipping the outcome from 0 to 1, that is closest to the factual. This is motivated by the minimality of actual causes in AC3 in Definition \ref{def:actual_cause}. Therefore, we first define a partial order on I/O behaviors.
\begin{definition}[Partial Order on Behaviors]
 \label{def:partial_order_components}
    For a factual behavior choice $\func: \inspace \mapsto \outspace$, we define a \emph{partial behavior order} $\preccurlyeq_\func$ on $(\inspace \mapsto \outspace)$ as 
    \begin{equation}
    \label{eq:partial_order_components}
        \begin{split}
            \func_1 \preccurlyeq_\func \func_2 &\Longleftrightarrow \forall x \in \inspace, \forall j \in \{1, \dots, \operatorname{dim}(\outspace)\},\\
            &(\func(x)[j] \leq \func_1(x)[j] \leq \func_2(x)[j]) \\
            \lor &(\func(x)[j] \geq \func_1(x)[j] \geq \func_2(x)[j])
        \end{split}
    \end{equation}
    where $[j]$ denotes the $j$-th dimension of a vector.
\end{definition}
In plain words, $\func_1 \preccurlyeq_\func \func_2$ iff $\func_1$ has closer outputs to the factual $\func$ than $\func_2$ does on all dimensions and on all inputs.}

\editn{However, one drawback of this discrete HP model and the partial order from Definition \ref{def:partial_order_components} is that we cannot tell the difference between two counterfactual I/O behaviors in terms of the sets of ``disagreeing'' nodes compared to the factual $\vec{v} = \discretize(\func_r)$} \edit{, which is something we need for reasoning about (AC3) in Definition \ref{def:actual_cause}.} \editn{For example, in Figure~\ref{fig:hp_model_ver_2}, assume that two counterfactuals $\func_1$ and $\func_2$ change the mapping on $\inspace_1$ from the factual $\outspace_2$ (as in $\func_r$) to $\outspace_3$ and $\outspace_4$, respectively. With this mapping change, both flip the outcome to 1. In the discretized HP model, the number of ``disagreeing'' nodes under both assignments is 1, but $\outspace_4$ may be further from the factual $\outspace_2$ than $\outspace_3$. 
In plain words, we cannot answer ``which one is more different from $\func_r$: $\func_1$ or $\func_2$?'' by simply comparing the two sets of nodes they modify from the factual value assignment.}

\subsection{Propositional HP Model}
\label{subsec:propositional_hp_model}


We now present a partial order, $\preccurlyeq_{\vec{v}}$, over node value differences and define how $\preccurlyeq_{\vec{v}}$ relates to partial order $\preccurlyeq_{\func}$ from Definition \ref{def:partial_order_components}. We then present an HP model construction technique, starting from the discretized HP model, that preserves $\preccurlyeq_{\vec{v}}$ and $\preccurlyeq_{\func}$.

\editn{For the new partial order, if the subset of nodes that differ between assignments $\vec{v}_2$ and factual $\vec{v}$ contains the subset of nodes that differ between $\vec{v_1}$ and $\vec{v}$, we say that $\vec{v}_1$ is closer to $\vec{v}$ than $\vec{v}_2$ does. We formulate this as another partial order, on value assignments of a subset of endogenous nodes $\hpmendoio \subseteq \hpmendo$. The value space of $\hpmendoio$ is denoted as $\hpmvarspaceendoio$.
\begin{definition}[Partial Order on HP Node Values]
\label{def:partial_order_node_values}
    On a set of HP model nodes $\mathcal{U}$ and its value space $\mathcal{V}$, given the factual value assignment $\vec{v} \in \mathcal{V}$, we can define a partial order $\preccurlyeq_{\vec{v}}$ that
    \begin{equation}
        \label{eq:partial_order_node_values}
        \vec{v}_1 \preccurlyeq_{\vec{v}} \vec{v}_2 \Longleftrightarrow \nodediff{\mathcal{U}}(\vec{v}, \vec{v}_1) \subseteq \nodediff{\mathcal{U}}(\vec{v}, \vec{v}_2), 
    \end{equation}
    where $\nodediff{\mathcal{U}}(\cdot, \cdot)$ denotes the subset of nodes in $\mathcal{U}$ that has different values by two assignments. Equality $=_{\vec{v}}$ holds iff $\nodediff{\mathcal{U}}(\vec{v}, \vec{v}_1) = \nodediff{\mathcal{U}}(\vec{v}, \vec{v}_2)$.
\end{definition}
}


With this partial order, we \edit{can determine} if an assignment $\vec{v}_1$ differs more from the factual $\vec{v}$ than assignment $\vec{v}_2$ does. 



\editn{Consequently, a larger change from the factual I/O behavior needs to be reflected in a larger set of differing nodes, i.e., the partial order $\preccurlyeq_{\vec{v}}$ needs to preserve the partial order $\preccurlyeq_{\func}$. We define partial order preservation as follows:}



\begin{definition}[Partial Order Preservation]
    \label{def:partial_order_preservation}
    Consider a representative \edit{behavior} space $\reconfuncs \subset (\inspace \mapsto \outspace)$ with a factual representative component $\func_r$ and the induced partial behavior order $\preccurlyeq_{\func_r}$. \edit{Consider also a subset of endogenous nodes, $\hpmendoio \subseteq \hpmendo$, with the value space of this subset $\hpmvarspaceendoio$.}
    An encoding of a representative I/O behavior onto the nodes $\hpmendoio$, $\encode: \reconfuncs \mapsto \hpmvarspaceendoio$, \emph{preserves partial order} iff \edit{$\encode(\func_r) = \vec{v}$}, and
\editn{\begin{equation}\label{eq:partial_order_preservation}
        \forall \func_1, \func_2 \in \reconfuncs, \func_1 \preccurlyeq_{\func_r} \func_2 \Longleftrightarrow \encode(\func_1) \preccurlyeq_{\vec{v}} \encode(\func_2)
    \end{equation}}
\end{definition}

\editn{Finally, we define a finer-grained HP model that preserves these two partial orders based on Definition \ref{def:partial_order_preservation}.}

\begin{figure}
    \centering
    \includegraphics[width=0.45\textwidth]{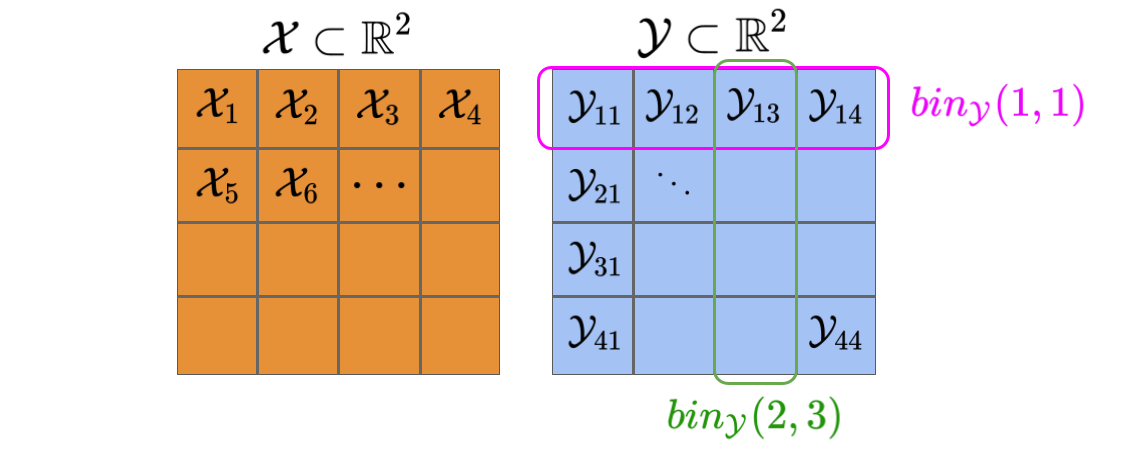}
    \caption{An example indexing of cells and output bins when both $\inspace$ and $\outspace$ are 2-dimensional.}
    \label{fig:indexing}
\end{figure}

\edit{First, we first define an output bin of $\outspace$. As illustrated in Figure \ref{fig:indexing}, if the output space is partitioned into $4 \times 4$ hypercube 
cells, we can index the cells from $\outspace_{11}$ to $\outspace_{44}$. We therefore have 4 bins along dimension 1 and 4 bins along dimension 2. Formally, if we index the $\text{dim}(\outspace) = d$-dimensional output cells as $\outspace_{k_1k_2\dots k_d}$, we have
\begin{definition}[Output Bins]
    The $k$-th output bin along dimension $j$ is $bin_\outspace(j, k) = \bigcup\{\outspace_{k_1k_2\dots k_d} \mid k_j = k\}$
\end{definition}
We denote the total number of bins along the $j$-th dimension as $n_j$, and therefore $k_j \in \{1, \dots, n_j\}$. Next, we can define the lower bound of $bin_\outspace(j, k)$ in $j$-th dimension as $lo(bin_\outspace(j, k))$. Notice that along a fixed dimension $j$, we have a total order of lower bounds of bins based on $k$.}

\begin{figure}
    \centering
    \includegraphics[width=0.5\textwidth]{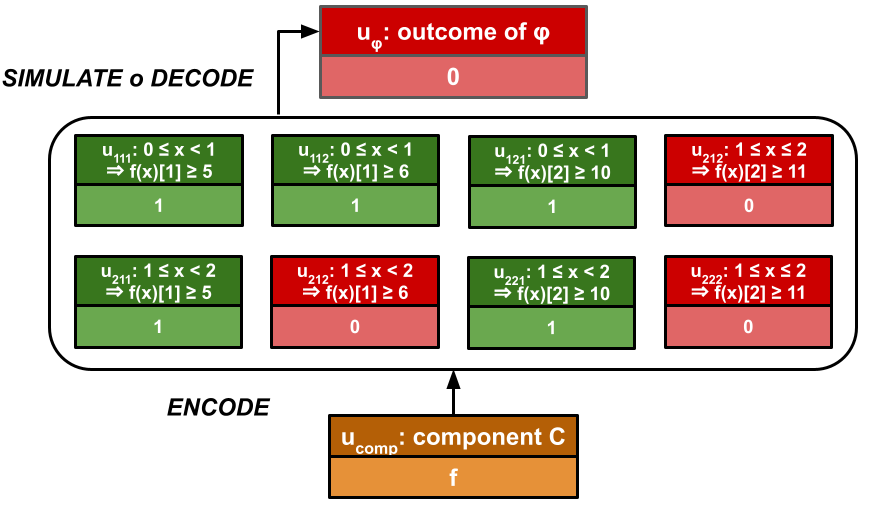}
    \caption{\edit{An example of the propositional HP model.} We use red and green colors to \edit{show the nodes with 0 and 1 values}.}
    \label{fig:hp_model_ver_3}
\end{figure}

\editn{We are now ready to} construct the \edit{final, \emph{propositional HP model}, created} by Algorithm \ref{alg:hp_model_construction}. 
\edit{In contrast to the discretized HP model}, now the endogenous nodes that encode the I/O behaviors ($\hpmendoio$ in the algorithm) \edit{take propositional values}. Node $u_{ijk}$ encodes \edit{whether for input $x \in \inspace_i$ the output cell for the $j$-th dimension is in at least the $k$-th output bin.} 
In the example illustrated in Figure \ref{fig:hp_model_ver_3}, we have 1-dimensional $\inspace$, with $\inspace_1 = [0, 1]$ and $\inspace_2 = [1, 2]$. We also have 2-dimensional $\outspace = ([5, 6] \cup [6, 7]) \times ([10, 11] \cup [11, 12])$. The example value shows that the representative function $\func_r$ of $\func$ maps all $x \in [0, 1]$ to the center of $[6, 7] \times [10, 11]$, and all $x \in [1, 2]$ to $[5, 6] \times [10, 11]$. \edit{Under this propositional HP model, the encoding \editn{$\encode: \hpmvarspaceendoio \mapsto \reconfuncs$} 
is simply evaluating the node propositions based on the component \editn{behavior}, and $\decode$ is first identifying the discretized mapping between cells, and then calling $\recon$. Notice that $\encode$ and $\decode$ are inverse to each other if we restrict the \editn{behavior} choices in $\reconfuncs$.}

\begin{algorithm}
\caption{Construct \edit{Propositional} HP Model}
\label{alg:hp_model_construction}
\SetKwInOut{Input}{Input}
\SetKwInOut{Output}{Output}
\SetKwRepeat{Do}{do}{while}
\Input{Input space $\inspace$ and output space $\outspace$ of component behaviors and their indexed cells, \edit{simulator $\simu$, encoder $\encode$, decoder $\decode$}}
\Output{HP model $\hpm = (\hpmendo, \hpmexo, \hpmvarspace, \hpmedges)$}
$\hpmexo \gets \{u_{\text{comp}}$\}\;
$\hpmendo \gets \{u_\property\}$\;
$\hpmvarspace(u_{\text{comp}}) \gets (\inspace \mapsto \outspace)$\;
$\hpmendoio \gets \emptyset$\;
\edit{$\hpmedges \gets \{\simu \circ \decode, \encode\}$\;}
\For{$i = 1, \dots, m$}{
    \For{$j = 1, \dots, d$}{
        \For{$k = 1, \dots, n_j$}{
            $u_{ijk} \gets (x \in \inspace_i \implies \func(x)[k] \geq lo(bin_\outspace(j, k))$\; 
            $\hpmvarspace(u_{ijk}) \gets \{0, 1\}$ \;
            $\hpmendoio \gets \hpmendoio \cup \{u_{ijk}\}$\;
        }
    }
}
$\hpmendo \gets \hpmendoio \cup \hpmendo$\;
\end{algorithm}

With $m$ input cells, $d$ output dimensions and $n_j$ bins along a dimension $j$, this HP model has $m\sum_{j=1}^dn_j + 2$ nodes, with $|\hpmendoio| = m\sum_{j=1}^dn_j$. However, the total number of value assignments on $\hpmendoio$ is not $2^{|\hpmendoio|}$, because some value assignments are not allowed. For example, we cannot let $f(x) \geq 5$ be false and $f(x) \geq 6$ be true. 
The total number of valid assignments is $(\prod_{j=1}^dn_j)^m = n^m$, the same as \edit{the discretized HP model}.


\begin{theorem}[Propositional Encoding Preserves Partial Order]
    \label{thm:partial_order_preservation}
The encoding $\encode: \reconfuncs \mapsto \hpmvarspaceendoio$ \editn{specified} by evaluating the proposition of every $u_{ijk} \in \hpmendoio$, \editn{as constructed in Algorithm \ref{alg:hp_model_construction},} preserves the partial order as defined in Definition \ref{def:partial_order_preservation}.
\end{theorem}

%% file: sections/5_diagnosis.tex
\section{The Causal Repair Algorithm}
\label{sec:diagnosis}

\subsection{Satisfactory Counterfactual Search}
\label{subsec:diagnosis_search}

Based on the \edit{constructed propositional} HP model, we look for an actual cause. The idea is to first search for a counterfactual $\valassignio'$ on $\hpmendoio$ that leads to a satisfactory outcome, i.e., $\sys(\func_r') \models \property$ and \edit{$\func_r' = \decode(\valassignio')$}. The subset of nodes with different value assignments between the factual $\valassignio$ and counterfactual $\valassignio'$ is not necessarily an actual cause yet (it may not be minimal as per condition AC3), and in the next subsection, we will look for a different $\valassignio^*$ that fulfills the actual cause conditions from this $\valassignio'$. Here, we focus on the search for $\valassignio'$.

\editn{Since a brute-force search on all $n^m$ possible value assignments takes exponential time, }
we leverage random sampling as follows in Algorithm~\ref{alg:counterfactual_sampling}.

\begin{algorithm}
\caption{Counterfactual Random Sampling}
\label{alg:counterfactual_sampling}
\SetKwInOut{Input}{Input}
\SetKwInOut{Output}{Output}
\SetKwRepeat{Do}{do}{while}
\Input{HP model $\hpm$, probability threshold $p$, significance level $\alpha \in [0, 1]$, simulator $\simu$, \edit{decoder $\decode$, probability distributions $\mathcal{D}_1$, $\mathcal{D}_2$}}
\Output{Either a counterfactual value assignment $\valassignio'$ on $\hpmendoio$ \edit{with $\simu(\decode(\valassignio')) = 1$, or a statement $\Pr_{\mathcal{D}_2}[\Pr_{\mathcal{D}_1}[\simu(\decode(\valassignio')) = 1] \leq p] \geq 1-\alpha$}}
$N \gets \lceil (1/p - 1)Q(1-\alpha/2)^2 \rceil$\;
\For{$1 \dots N$}{
    $\valassignio'\gets$ uniform sampling from all $n^m$ settings\;
    \edit{$\property \gets \simu(\decode(\valassignio'))$\;}
    \If{$\property$}{
        Return $\valassignio'$ \;
    }
}
\edit{Return $\Pr_{\mathcal{D}_2}[\Pr_{\mathcal{D}_1}[\simu(\decode(\valassignio')) = 1] \leq p] \geq 1-\alpha$\;}
\end{algorithm}
The distribution $\mathcal{D}_1$ is a distribution on different value assignments in the value space $\hpmvarspaceendoio$. For example, it can be a uniform distribution on the propositional node values. Distribution $\mathcal{D}_2$ is the induced distribution on a sampled value assignment's success rate, \editn{i.e., $p = \Pr_{\mathcal{D}_1}[\simu(\decode{(\valassignio')} = 1]$}
, which is treated as another random variable. In line 1, the function $Q(\cdot)$ means the quantile on standard normal distribution. Starting from line 2 to 8, we \edit{uniformly} sample different value assignments $\valassignio'$
until we find one $\valassignio'$ that produces satisfactory $\property$ by $\simu
$ \edit{or we reach a maximum number of samples.} If we fail to find a satisfactory value assignment, we report 
that the probability of finding a satisfactory $\valassignio'$ with uniform sampling, 
i.e., the portion of satisfactory $\valassignio'$ in the entire search space, is at most $p$ with confidence $1-\alpha$ at line 9.
We show our probabilistic failure statement at line 9 holds with the following theorem.

\begin{theorem}[Probabilistic Guarantee on Search Failure]
\label{thm:probabilistic_guarantee}
Given an HP model constructed by Algorithm \ref{alg:hp_model_construction},
a probability threshold $p \in [0, 1]$ and a confidence $1-\alpha \in [0, 1]$, the final statement at line 9 of Algorithm \ref{alg:counterfactual_sampling} holds.
\end{theorem}


\looseness=-1
Upon $N$ consecutive failures, this search algorithm states that with some confidence a counterfactual \edit{assignment that flips the outcome} is unlikely to be found.
This suggests that the actual cause for the run-time property violation lies elsewhere: possibly the environment, \editn{the behaviors of } another component, or the suspected component together with another component --- but not \editn{the I/O behaviors of} this component alone. This confidence is based on a substantial number of samples without finding a successful counterfactual. For example, for $p = 0.001$ and $\alpha = 0.05$, one would need to uniformly sample at least $N=3838$ failed counterfactuals in a row.

\subsection{Node Value Interpolation for Actual Cause}
\label{subsec:diagnosis_interpolation}

Suppose we have successfully obtained a satisfactory $\valassignio'$ on $\hpmendoio$, from Algorithm \ref{alg:counterfactual_sampling}. The 
final step is to find a counterfactual $\valassignio^*$ \edit{that can flip the outcome and is as close to the factual $\valassignio$ as possible, starting from $\valassignio'$.
This step is required to satisfy the minimiality condition (AC3) of Definition \ref{def:actual_cause}.}


\editn{We therefore perform a deterministic interpolation between $\valassignio'$ and $\valassignio$ in Algorithm \ref{alg:interpolation}, 
which starts from the found satisfactory counterfactual assignment $\valassignio'$. In this algorithm, the counterfactual value assignment steps towards the factual $\valassignio$ by flipping the differing value assignments one-by-one. This procedure continues until there are no nodes it can flip while still satisfying $\property$.}
\edit{Because} the total number of nodes in $\hpmendoio$ is $m\sum_{j=1}^d n_j$, Algorithm \ref{alg:interpolation} is guaranteed to output an actual cause with a satisfactory counterfactual within $O(m\sum_{j=1}^d n_j)$ time. The complexity is linear in terms of $m$ and $n$, allowing for efficient computation.

\begin{figure}
    \centering
    \includegraphics[width=0.3\textwidth]{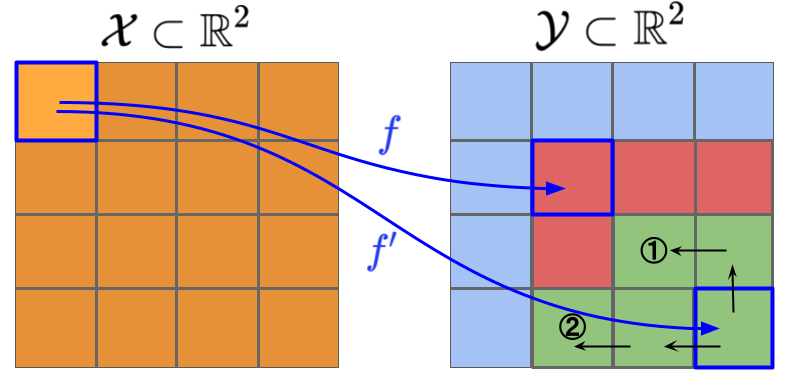}
    \caption{An example of finding actual cause by interpolation, with two-dimensional $\inspace$ and $\outspace$. Red cells in $\outspace$ denote property violation, green denotes satisfaction, and blue denotes cells where satisfaction is not relevant to the example.}
    \label{fig:interpolation_demo}
\end{figure}

We visualize the process of searching for $\valassignio^*$ as Figure \ref{fig:interpolation_demo}. Without loss of generality, consider only the top left input cell. The factual \editn{behavior} \edit{$\func = \decode(\valassignio)$} maps inputs in this cell to some output cell and this behavior produces a property violation. The sampling by Algorithm \ref{alg:counterfactual_sampling} finds a satisfactory counterfactual \edit{$\func' = \decode(\valassignio')$}, which is an alternative mapping. Next, the interpolation Algorithm \ref{alg:interpolation} 
flips the nodes disagreed by $\valassignio$ and $\valassignio'$ one-by-one towards $\valassignio$. This flipping is equivalent to stepping through the output cells one by one. \edit{Eventually, the algorithm reaches a node assignment where any steps towards $\func$ result in $\property$ becoming violated, at which point it returns the current node assignment.}
Depending on the dimensions, we can take different paths in stepping, i.e. we can end up either in cell (1) or cell (2), from interpolating in the vertical or horizontal dimension first, respectively. Mapping the input cell to either of these two cells represents a valid I/O behavior of $\solfunc$.

\begin{algorithm}
\caption{Actual Cause Search by \edit{Incremental} Interpolation}
\label{alg:interpolation}
\SetKwInOut{Input}{Input}
\SetKwInOut{Output}{Output}
\SetKwRepeat{Do}{do}{while}
\Input{HP model $\hpm$, factual $\valassignio$, satisfactory counterfactual $\valassignio'$, simulate function $\simu$}
\Output{Satisfactory counterfactual $\valassignio^*$ such that its difference from $\valassignio$ is an actual cause as per $\hpm$}
$\valassignio^* \gets \valassignio'$\;
$\nodediffio^* \gets \nodediffio(\valassignio, \valassignio')$\;
\For{$i = 1, \dots, m$}{
    \For{$j = 1, \dots, d$}{
        \For{$k = 1, \dots, n_d$}{
            \If{$u_{ijk} \notin \nodediffio^*$}{
                Continue\;
            }
            \editn{
            $\vec{v}_{\text{temp}} \gets$ $\valassignio^*$ with value assignment on $u_{ijk}$ the negation as in $\valassignio^*$\;}
            \editn{\If{$\simu(\decode(\vec{v}_{\text{temp}})) = $ 1}{
                $\valassignio^* \gets \vec{v}_{\text{temp}}$\;
                $\nodediffio^* \gets \nodediffio^* \setminus \{u_{ijk}\}$
            }}
        }
    }
}
\end{algorithm}



\edit{Notice that Algorithm \ref{alg:interpolation} incrementally steps towards the factual $\valassignio$ from $\valassignio'$ by flipping nodes one-by-one. A more efficient variant is to do a \emph{binary search} between these two value assignments. We denote these two approaches as incremental interpolation and binary search interpolation, respectively. Both are evaluated in Section \ref{sec:expriments}.}


Next, we show that the output is indeed an actual cause based on HP model $\hpm$ in Theorem~\ref{thm:output_actual_cause}.

\begin{theorem}[Output is Actual Cause]
    \label{thm:output_actual_cause}
    \editn{Let HP model $\hpm$ constructed in Algorithm~\ref{alg:hp_model_construction} be given with factual node value assignment $\valassignio$. Let $\valassignio'$ be a counterfactual node value assignment from Algorithm \ref{alg:counterfactual_sampling}. }
    The node values on $\hpmendoio$ \edit{where assignments} $\valassignio$ and $\valassignio^*$ \edit{disagree in} Algorithm~\ref{alg:interpolation} are an \emph{actual cause} of $\sys(\func) \not\models \property$ as per $\hpm$ \edit{constructed in Algorithm~\ref{alg:hp_model_construction}.}
\end{theorem}


%% file: sections/6_experiments.tex
\section{Experimental Evaluation}
\label{sec:expriments}

\subsection{Setup}
\label{subsec:expr_setup}


\begin{figure*}
    \centering
    \includegraphics[width=\textwidth]{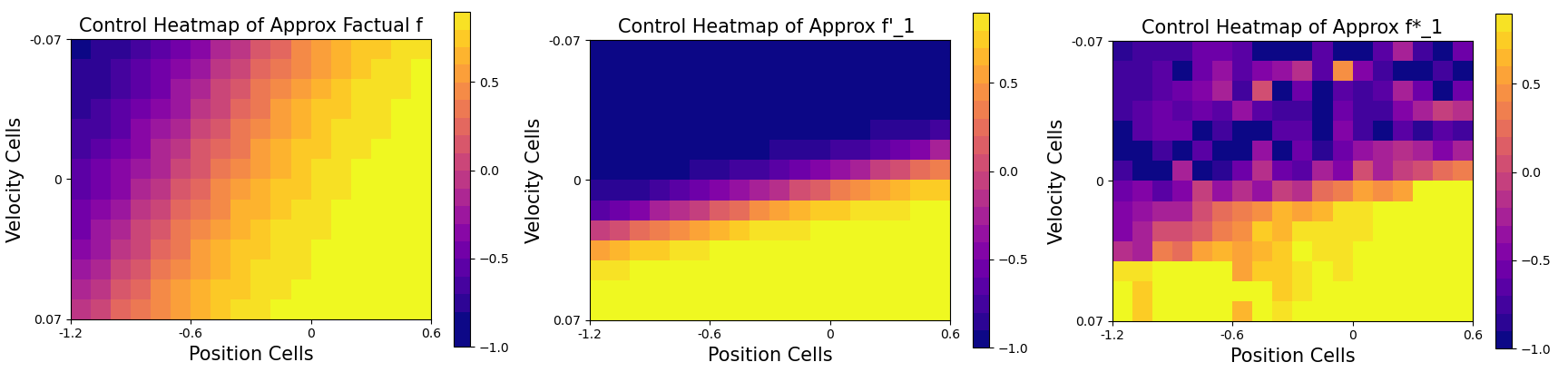}
    \caption{Control functions, approximated by mapping between cells, visualized as heatmaps on the position-velocity space. The functions from left to right are the factual $\func$, the searched counterfactual $\func_1'$ and interpolated counterfactual $\solfunc_1$ between the first two.}
    \label{fig:heatmaps}
\end{figure*}

We test our diagnosis on the mountain car from the OpenAI Gym~\cite{brockman2016openai},
\edit{which has become a common benchmark for learning-enabled CPS~\cite{ivanov_verisig_2021,ruchkin_confidence_2022}}. The car starts in the valley between two mountains \edit{with} the task is to drive it to the top of the right mountain within a deadline. The system has two one-dimensional state variables: position and velocity, and a one-dimensional control signal, as follows.
\begin{equation}
    \label{eq:mc_system}
    \begin{split}
        &pos(t+1) = pos(t) + vel(t)\\
        &vel(t+1) = vel(t) + 0.0015 ctrl(t) - z  \operatorname{cos}{(3 pos(t))}\\
        &ctrl(t) = \func(pos(t), vel(t))
    \end{split}
\end{equation}
where $z=0.0025$ is the steepness of the hill. The variables are bounded, such that $pos(t) \in [-1.2, 0.6]$, $vel(t) \in [-0.07, 0.07]$ and $ctrl(t) \in [-1, 1]$. The initial condition is $(pos(0), vel(0)) = (-0.5, 0)$, i.e., staying still at the bottom of the valley. The controller is \edit{a learning-based}  function $\func$, \edit{for which we use} a pre-trained deep neural network. The run-time property can be specified as an STL \cite{maler2004monitoring} formula 
\begin{equation}
    \label{eq:mc_property}
    \property = F_{t \leq 110}(pos(t) \geq 0.45),
\end{equation}
i.e., reaching $pos(t) \geq 0.45$ before the deadline of $t=110$. \edit{Due to its} limited  power, the car \edit{cannot} reach its goal \edit{directly,} and  the challenge is to \edit{first climb the left mountain} to gain enough momentum.

As the controller function $\func$, we use a pre-trained rectangular deep neural network with shape $8 \times 16$ with sigmoid activations. The run-time property $\property$ is violated \editn{on initial state $(pos(0), vel(0)) = (-0.5, 0)$}
, and we suspect it is caused by this learned controller. Therefore, we run our diagnosis step-by-step to find an actual cause and observe its effect on system repair. All experiments are run on a single core of Intel Xeon Gold 6148 CPU @ 2.40GHz.

\subsection{Results}
\label{subsec:expr_results}

Based on the setup, the input space and output space of the suspected controller are $\inspace = [-0.6, 1.2] \times [-0.07, 0.07]$ and $\outspace = [-1, 1]$. We first run Algorithm \ref{alg:io_partition} and find that the cell width is of size $0.1 \times 0.01$ on $\inspace$ and $0.1$ on $\outspace$. In other words, position is split into $(1.2 - (-0.6)) / 0.1 = 18$ equally sized intervals, velocity into $(0.07 - (-0.07)) / 0.01 = 14$ intervals and control into $(1 - (-1)) / 0.1 = 20$ intervals. Therefore, there are $18 \times 14 = 252$ input cells and $20$ output cells.

By using Algorithm \ref{alg:hp_model_construction}, we constructed an HP model, with every $u_{ik} \in \hpmendoio$ representing a Boolean statement ``on this input cell $\inspace_i$ of (position, velocity), the output control is at least in output bin $bin_\outspace(1, k)$''. We do not have $j$ here because the output space is one-dimensional. Then, using Algorithm \ref{alg:counterfactual_sampling}, we found \edit{three} counterfactuals 
\editn{$\valassignio^1$, $\valassignio^2$, and $\valassignio^3$}
that lead to the satisfaction of $\property$. We then applied Algorithm \ref{alg:interpolation} to find the minimal modification needed from factual $\valassignio$ to these 3 counterfactuals, with the interpolated counterfactuals denoted as $\valassignio^{1*}$, $\valassignio^{2*}$ and $\valassignio^{3*}$, respectively. 

Figure \ref{fig:heatmaps} shows the control functions $\func$, \editn{$\func'_1 = \decode(\valassignio^1)$, and $\solfunc_1 = \decode(\valassignio^{1*})$ }
as heatmaps. There are 153 (out of 252) input cells that map to a different output \edit{cell between} $\func$ and $\solfunc_1$. Consequently, our diagnosis pipeline concludes \edit{that} the actual cause is \emph{``the control signals on these 153 input cells given by the factual controller $\func$''}, with a corresponding repair \emph{``had these 153 cells been mapped by $\solfunc_1$ instead of $\func$, the system would have satisfied $\property$''}. 

\editn{After we found a counterfactual behavior, we used  both the incremental and binary search interpolations. The computation time is listed in Table~\ref{tab:interpolation_time}. We can see the time overhead is predominantly from running the simulator, and that interpolation with binary search does reduce this overhead.
\begin{table*}[]
    \centering
    \begin{tabular}{|c|c|c|c|c|}
        \hline
        Interpolation & Total time (s) & Simulator time (s) & Stepping time (s) & \# of operations \\
        \hline
        Incremental & 9148.77 & 9148.76 & 0.003 & 1231\\
        Binary & 4965.03 & 4965.02 & 0.001 & 880\\
        \hline
    \end{tabular}
    \caption{Computation time of interpolation from counterfactual $\func_1'$ to factual $\func$. An ``operation'' is a combination of executing \simu~ and stepping the counterfactual towards the factual behavior.}
    \label{tab:interpolation_time}
\end{table*}}

\begin{figure}
    \centering
\includegraphics[width=0.5\textwidth]{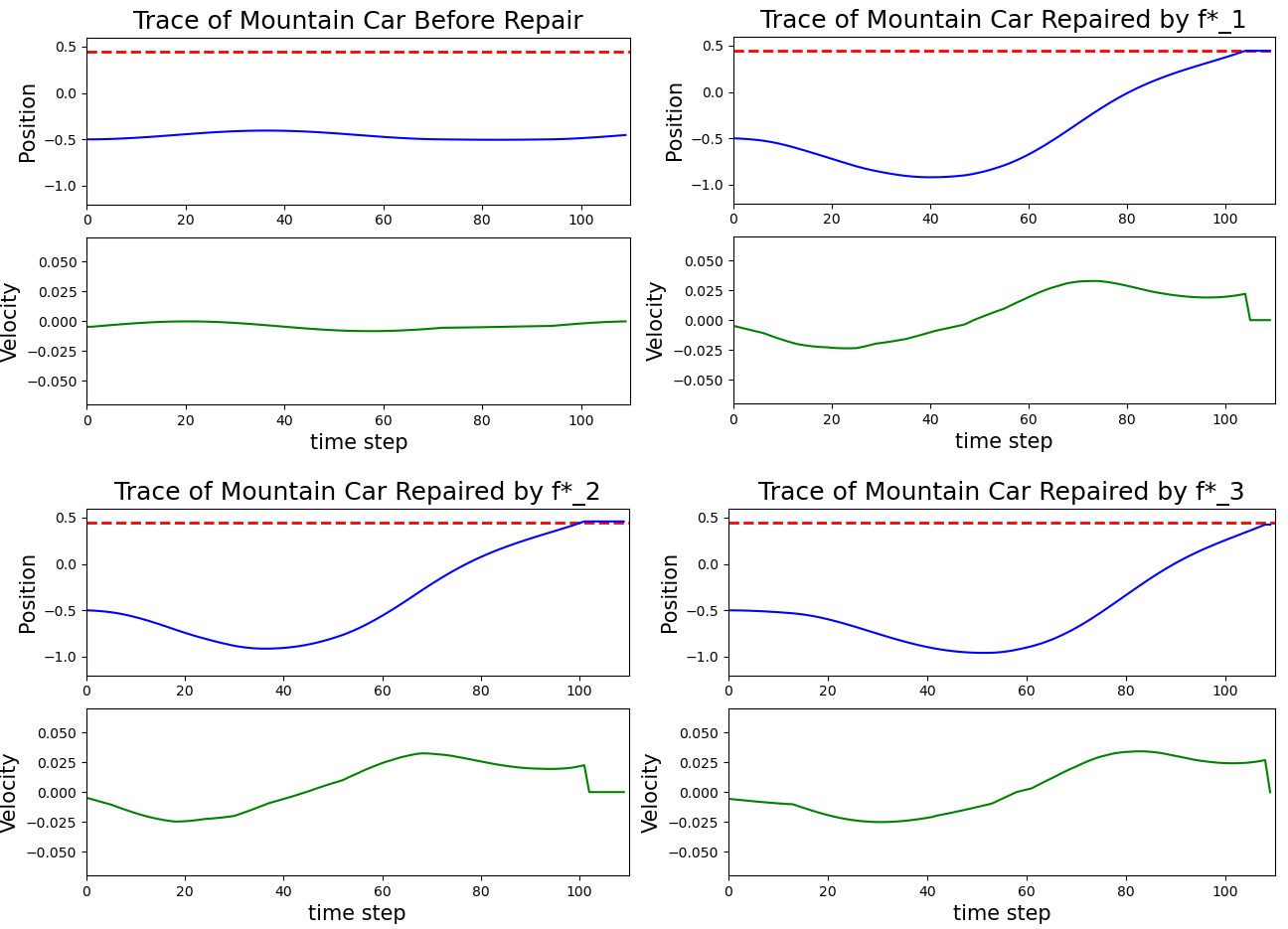}
    \caption{Mountain car traces: one before repair and three causally repaired. The red dashed line means $pos = 0.45$, i.e., the target. The car stops immediately after target is reached.}
    \label{fig:mc_traces}
\end{figure}

To validate that these modifications indeed repair the system, we replace the controller with interpolated control functions $\solfunc_1$, $\solfunc_2$ and $\solfunc_3$. The re-runs give satisfactory results as shown in Figure \ref{fig:mc_traces}, \edit{showing} the utility of our  causal diagnosis.

%% file: sections/7_discussion.tex
\section{Discussion and Conclusion}
\label{sec:discussion}

Our diagnosis \edit{and repair} pipeline
has several advantages. First, we are able to leverage domain knowledge of the system\edit{, such as how to simulate the outcome of the given STL property,}
and well-defined actual causality models to reason about the causes. 
Second, the actual cause we identify 
reflects a minimal subset of the component's I/O behaviors \edit{that cause the problem under our defined partial order}. 
Therefore, we avoid \edit{changing} the complex and possibly unavailable internal structure of components and blaming unrelated, ``innocent'' behaviors. Finally, as the experiments demonstrate, our approach produces \edit{practically} useful repair suggestions. 

Limitations do exist in our current pipeline, so future research is required. First, the HP model construction and component behavior encoding in Section \ref{subsec:propositional_hp_model} is only one way of preserving the partial order \edit{on behaviors}. There may be more effective ways with fewer HP model nodes and therefore smaller memory and time overhead. Second, the search algorithm in Section \ref{subsec:diagnosis_search} \edit{draws} uniformly random samples from an exponentially large search space, and even if we are confident to report the satisfactory portion is relatively small and hence the search fails, that portion may still be large in an absolute sense. For example, a fraction of $10^{-6}$ on a space size of $10^{20}$ is still \edit{very large}. Therefore, there is potential for making the search algorithm smarter, such as leveraging prior knowledge of the STL property. Moreover, we \edit{perform a repair on only one} initial condition, which may \edit{violate the property on the others}. 

%% file: sections/8_conclusion.tex

To summarize, this paper uses a novel construct of the Halpern-Pearl model to search for an actual cause for a run-time CPS property violation and provide a repair, without \edit{changing} complex internal structures of LECs. Our diagnostic pipeline \editn{consists of two steps: (1) construct the propositional HP model and (2) search for different variable assignments on the model until we reach an actual cause.}
Our \edit{causal pipeline provides repair guarantees and can generate  useful fixes to deep neural network controllers}.

%% file: sections/acknowledgement.tex
\editcr{
\section*{Acknowledgement}
This work was supported in part by ARO W911NF-20-1-0080 and AFRL and DARPA FA8750-18-C-0090. Any opinions, findings and conclusions or recommendations expressed in this material are those of the authors and do not necessarily reflect the views of the Air Force Research Laboratory (AFRL), the Army Research Office (ARO), the Defense Advanced Research Projects Agency (DARPA), or the Department of Defense, or the United States Government.
}

%% file: sections/references.tex

%% file: sections/Appendix_A_proofs.tex
\section{}
\label{sec:appendix_proofs}


In this appendix, we provide the proofs of the following theorems from paper ``Causal Repair of Learning-Enabled Cyber-Physical Systems'': Theorem \ref{thm:alg1_termination} from Section \ref{subsec:discrete_hp_model}, Theorem \ref{thm:partial_order_preservation} from Section \ref{subsec:propositional_hp_model}, Theorem \ref{thm:probabilistic_guarantee} from Section \ref{subsec:diagnosis_search}, and Theorem \ref{thm:output_actual_cause} from Section \ref{subsec:diagnosis_interpolation}. 

First, we prove that Algorithm \ref{alg:io_partition} is guaranteed to terminate.

\begin{customthm}{1}[Termination of Discretization]
    \label{thm:alg1_termination}
    Algorithm \ref{alg:io_partition} is guaranteed to terminate.
\end{customthm}

\begin{proof}
Within an input cell $\inspace_i$ of width $\Delta_x$, for any two $x_1$, $x_2$ inside this cell, we have
\begin{equation}
    ||x_1 - x_2||_{d_\inspace} \leq \sqrt{\text{dim}(\inspace)}\Delta_x,
\end{equation}
i.e., any two points within this cell must be smaller than the main diagonal length of cell $\inspace_i$.

Next, by Assumption \ref{assum:lipchitiz}, the behavior of factual component $\func$ is Lipschitz-continuous, so $\exists$ Lipchitiz constant $c > 0$, that
\begin{equation}
    \forall x_1, x_2 \in \inspace, ||\func(x_1) - \func(x_2)||_{d_\outspace} \leq c||x_1 - x_2||_{d_\inspace}.
\end{equation}
 Therefore, when these inputs are mapped into the output space by $\func$, we have
\begin{equation}
    ||f(x_1) - f(x_2)||_{d_\outspace} \leq c\sqrt{\text{dim}(\inspace)}\Delta_x,
\end{equation}
which means there exists a cubic output cell $\outspace_j$ of width $\Delta_y \geq c\left(\sqrt{\text{dim}(\inspace)}/\sqrt{\text{dim}(\outspace)}\right)\Delta_x$ such that every input from $\inspace_i$ can be mapped to $\outspace_j$.

In this cell partition
, the maximal difference between any output $y \in \outspace_j$ and its cell center $\text{center}(\outspace_j)$ is bounded by half of the main diagonal length, i.e.,
\begin{equation}
    ||y - \text{center}(\outspace_j)||_{d_\outspace} \leq \sqrt{\text{dim}(\outspace)}\Delta_y.
\end{equation}
Since every output is represented by its cell center, by Definition \ref{def:component_distance}, 
\begin{equation}
    ||\func - \func_r||_{d_\func} \leq \sqrt{\text{dim}(\outspace)}\Delta_y.
\end{equation}
Recall that Assumption \ref{assum:robustness} states that the outcome of $\simu$ maintains the same when two functions has a separation smaller than $\epsilon$. Therefore, we shrink $\Delta_x$ (in lines 7-10) until $ \sqrt{\text{dim}(\outspace)}\Delta_y \leq \epsilon$, so that the representative function $\func_r$ leads to the same outcome, i.e. runtime property violation, as the factual behavior $\func$.


Therefore, we are able to find the representative function $\func_r$ constructed at line 12. 
\end{proof}


Next, we prove that the propositional HP model constructed by Algorithm \ref{alg:hp_model_construction} preserves the partial order from Definition \ref{def:partial_order_preservation}.

\begin{customthm}{2}[Encoding Preserves Partial Order]
The encoding $\encode: \reconfuncs \mapsto \hpmvarspaceendoio$ specified by evaluating the proposition of every $u_{ijk} \in \hpmendoio$,as constructed in Algorithm \ref{alg:hp_model_construction}, preserves the partial order as defined in Definition \ref{def:partial_order_preservation}.
\end{customthm}

\begin{proof}
We start by proving the forward direction $\Longrightarrow$ in Definition \ref{def:partial_order_preservation}. Suppose we have two arbitrary behavior choices choices $\func_1, \func_2 \in \reconfuncs$. The factual behavior is represented by $\func_r$ and we have $\func_1 \preccurlyeq_{\func_r} \func_2$.

By Definition \ref{def:partial_order_components}, this partial order $\preccurlyeq_{\func_r}$ means that $\func_1$ has an output in between that of $\func_r$ and $\func_2$ in every output dimension, i.e.,
\begin{equation}
    \label{eq:proof_component_partial_order}
    \begin{split}
        &\forall x \in \inspace, \forall j \in \{1, \dots, \text{dim}(\outspace)\},\\
        &(\func_r(x)[j] \leq \func_1(x)[j] \leq \func_2(x)[j]) \\
        &\lor (\func_r(x)[j] \geq \func_1(x)[j] \geq \func_2(x)[j]) 
    \end{split}
\end{equation}

We pick an arbitrary input $x \in$ input cell $\inspace_i$, and an arbitrary output dimension $j$. By Equation \ref{eq:proof_component_partial_order}, we know that the output bins that $\func_r(x)$, $\func_1(x)$ and $\func_2(x)$ in dimension $j$ must be in either non-decreasing, or non-increasing order. Without loss of generality, we consider non-decreasing order here. That is,
\begin{equation}
    \begin{split}
        &\func_r(x) \in bin_{\outspace}(j, k_r)\\
        &\func_1(x) \in bin_{\outspace}(j, k_1)\\
        &\func_2(x) \in bin_{\outspace}(j, k_2)\\
        &k_r \leq k_1 \leq k_3
    \end{split}
\end{equation}
Therefore, we have $lo(bin_{\outspace}(j, k_r)) \leq lo(bin_{\outspace}(j, k_1)) \leq lo(bin_{\outspace}(j, k_2))$. Now we consider the encoded node values on the subset of nodes $\hpmendoio(i, j) \subseteq \hpmendoio$, which are the nodes that encodes the behaviors of input cell $\inspace_i$ on dimension $j$, i.e., $\hpmendoio(i, j) = \{u_{ijk} \mid k = 1, \dots, n_j\}$. The value assignments on $\hpmendoio(i, j)$ given by the three behavior choices are denoted as $\vec{v}_r(i,j)$, $\vec{v}_1(i,j)$ and $\vec{v}_2(i,j) \in \hpmvarspaceendoio(i, j)$, respectively. 

Based on the order $lo(bin_{\outspace}(j, k_r)) \leq lo(bin_{\outspace}(j, k_1)) \leq lo(bin_{\outspace}(j, k_2))$, evaluating the node propositions, i.e., the encoding function $\encode()$, results in
\begin{equation}
    \label{eq:value_assignments}
    \begin{split}
        &\vec{v}_r(i,j) = \underbrace{111...1}_{k_r}\underbrace{000...0}_{n_j - k_r}\\
        &\vec{v}_1(i,j) = \underbrace{111...11}_{k_1}\underbrace{00...0}_{n_j - k_1}\\
        &\vec{v}_2(i,j) = \underbrace{111...111}_{k_2}\underbrace{0...0}_{n_j - k_2}
    \end{split}
\end{equation}
From the evaluated assignments, we can see the two susbets of differing nodes between $\vec{v}_r$ and $\vec{v}_1$ and between $\vec{v}_r$ and $\vec{v}_2$ are
\begin{equation}
    \begin{split}
        \text{diff}_{\hpmendoio}(i,j)(\vec{v}_r, \vec{v}_1) = \{u_{ijk} \mid k_r+1 \leq k \leq k_1\}\\
    \text{diff}_{\hpmendoio}(i,j)(\vec{v}_r, \vec{v}_2) = \{u_{ijk} \mid k_r+1 \leq k \leq k_2\}
    \end{split}
\end{equation}
And we can see
\begin{equation}
    \text{diff}_{\hpmendoio}(i,j)(\vec{v}_r, \vec{v}_1) \subseteq \text{diff}_{\hpmendoio}(i,j)(\vec{v}_r, \vec{v}_2)
\end{equation}
The same set containment holds when we use $\geq$, i.e. non-increasing order.

Notice that the three integers $k_r$, $k_1$ and $k_2$ denote the number of nodes flipped to 1 in $\hpmendoio(i, j)$ by these 3 behavior choices. Therefore, we can as well denote them as $k_r(i, j)$, $k_1(i, j)$ and $k_2(i, j)$. Since Equation \ref{eq:proof_component_partial_order} applies for all input cells $\inspace_i$ and output dimensions $j$, we have $k_r(i, j) \leq k_1(i, j) \leq k_2(i, j)$ (or $\geq$) for each pair of $i$ and $j$. Therefore, on all $i, j$, $\text{diff}_{\hpmendoio}(i,j)(\vec{v}_r, \vec{v}_1) \subseteq \text{diff}_{\hpmendoio}(i,j)(\vec{v}_r, \vec{v}_2)$ holds. Consequently,
\begin{equation}
    \begin{split}
        &\text{diff}_{\hpmendoio}(\vec{v}_r, \vec{v}_1) = \bigcup_{i,j}\text{diff}_{\hpmendoio}(i,j)(\vec{v}_r, \vec{v}_1) \\
        \subseteq &\bigcup_{i,j}\text{diff}_{\hpmendoio}(i,j)(\vec{v}_r, \vec{v}_2) = \text{diff}_{\hpmendoio}(\vec{v}_r, \vec{v}_2)
    \end{split}
\end{equation}
By Definition \ref{def:partial_order_node_values}, we have $\vec{v}_1 \preccurlyeq_{\vec{v}_r} \vec{v}_2$, and the forward direction $\Longrightarrow$ is proved.

To prove the backward direction $\Longleftarrow$, we can use a symmetrical approach as above. That is, we still split $\hpmendoio$ into multiple $\hpmendoio(i, j)$. Since we assume $\text{diff}_{\hpmendoio}(\vec{v}_r, \vec{v}_1) \subseteq \text{diff}_{\hpmendoio}(\vec{v}_r, \vec{v}_2)$, we must have $\forall i, j, \text{diff}_{\hpmendoio}(i,j)(\vec{v}_r, \vec{v}_1) \subseteq \text{diff}_{\hpmendoio}(i,j)(\vec{v}_r, \vec{v}_2)$. Consequently, we can show that $\func_1(x)[j] \leq \func_2(x)[j]$ (or $\geq$) on arbitrary input space $x \in \inspace_i$ and output dimension $j$
based on the construction in Algorithm \ref{alg:hp_model_construction}, and therefore $\func_1 \preccurlyeq_{\func_r} \func_2$. This encoding indeed preserves partial order in both directions.

\end{proof}


Next, we prove the validity of the probabilistic statement reported by Algorithm \ref{alg:counterfactual_sampling} when it cannot find a suitable counter example.


\begin{customthm}{3}[Probabilistic Guarantee on Search Failure]
Given an HP model constructed by Algorithm \ref{alg:hp_model_construction}, 
a probability threshold $p \in [0, 1]$ and a confidence $1-\alpha \in [0, 1]$, the final statement at line 9 of Algorithm \ref{alg:counterfactual_sampling} holds.
\end{customthm}

\begin{proof}
Let $\mathcal{D}_1$ denote the uniform distribution on value assignments on $\hpmendoio$, i.e., all propositional assignments on the binary nodes.
Consider random variable $\vec{v} \sim \mathcal{D}_1$. Let $s=\simu(\decode(\vec{v}))$ denote whether $\sys$ is safe or not when component $\mathcal{C}$ is replaced by a counter factual $\func'=\decode(\vec{v})$. Note that $s$ is a Bernoulli random variable, as it has value $1$ when $\sys$ is safe and $0$ otherwise and that $Pr_{\mathcal{D}_1}[s = 1]$ is the number of safe assignments divided by the total number of feasible assignments $n^m$.



For a Bernoulli distribution, the number of successes, denoted $N_s$, from $N$ sequential samples follows a Binomial distribution. If $N_s = 0$ we can apply the Wilson score interval \cite{newcombe1998interval} to get an upper bound on the success probability of the Bernoulli random variable:
\begin{equation}
    \label{eq:wilson_interval}
    \begin{split}
      z &= Q(1-\frac{\alpha}{2})\\
      \Pr[\text{success}] &\in \frac{N_s + \frac{z^2}{2}}{N + z^2} \pm \frac{z}{N + z^2}\sqrt{\frac{N_s(N-N_s)}{N} + \frac{z^2}{4}}\\
      &= \frac{z^2}{2N + 2z^2} \pm \frac{z^2}{2N + 2z^2}\\
      &= [0, \frac{z^2}{N + z^2}]
    \end{split}
\end{equation}

Equation \eqref{eq:wilson_interval} means that, if we do not see any success in $N$ Bernoulli trials, then with $1-\alpha$ confidence we estimate that the probability of a success in one trial is less than $z^2/(N+z^2)$, with $z$ being the $(1-\alpha/2)$-quantile of standard normal distribution.

Therefore, given an upper-bound probability estimation $p$, we can compute $N$ in terms of $p$ and $\alpha$.
\begin{equation}
    \label{eq:compute_N}
    \begin{split}
        &p = \frac{z^2}{(N+z^2)}\\
        \implies &N = (\frac{1}{p} - 1)z^2 = (\frac{1}{p} - 1)Q(1-\frac{\alpha}{2})^2
    \end{split}
\end{equation}
Therefore, if we sample at least $(1/p - 1)Q(1-\alpha/2)^2$ value assignments without any success, we can make the claim in line 9, supported by Wilson score interval. Notice that this minimal number of sampling is met at line 1. Consequently, the statement at line 9 holds.
\end{proof}


In the end, we prove the output of Algorithm \ref{alg:interpolation} is an actual cause given HP model $\hpm$.

\begin{customthm}{4}[Output is Actual Cause]
    Let HP model $\hpm$ constructed in Algorithm~\ref{alg:hp_model_construction} be given with factual node value assignment $\valassignio$. Let $\valassignio'$ be a counterfactual node value assignment from Algorithm \ref{alg:counterfactual_sampling}.
    The node values on $\hpmendoio$ where assignments $\valassignio$ and $\valassignio^*$ disagree in Algorithm~\ref{alg:interpolation} are an \emph{actual cause} of $\sys(\func) \not\models \property$ as per $\hpm$ constructed in Algorithm~\ref{alg:hp_model_construction}.
\end{customthm}

\begin{proof}

First, the encoding of the factual function, $\valassignio$ gives property violation $\neg \property$ based on Algorithm \ref{alg:io_partition} and \ref{alg:hp_model_construction}. Therefore, AC1 holds.

Second, if we partition $\hpmendo$ into $\mathcal{U}_1 = \nodediffio^* \cup \{u_\property\}$, $\mathcal{U}_2 = \emptyset$ and  $\mathcal{U}_3 = \hpmendo \setminus \mathcal{U}_1$, we can check this partition on AC2. The subset $\mathcal{U}_3$ does not have counterfactual values in these assignments so there is no difference between the counterfactuals in AC2(a) and AC2(b). In fact, the existence of this partition means AC2 holds.

Finally, there does not exist a subset $\text{diff}_{\hpmendoio}^{**} \subset \text{diff}_{\hpmendoio}^*$ such that by using counterfactual values on this smaller subset only gives satisfaction of $\property$, as then Algorithm \ref{alg:counterfactual_sampling} would have output the node value assignment corresponding to that counterfactual instead of outputting $\valassignio'$. 

Since all the three conditions hold, the factual value assignments on the disagreeing nodes, i.e., $\text{diff}_{\hpmendoio}^*$, is an actual cause of the property violation.
\end{proof}